\documentclass{article}

\usepackage{amsmath, amssymb, tensor, comment, graphicx, calc, enumitem, soul, natbib}

\setcitestyle{authoryear,open={(},close={)}} 

\newtheorem{proposition}{Proposition}
\newtheorem{definition}{Definition}
\newtheorem{theorem}{Theorem}

\title{Torsion in the Classical Spacetime Context}
\author{Helen Meskhidze and James Owen Weatherall}
\date{}

\begin{document}
\maketitle

\begin{centering}
Department of Logic and Philosophy of Science \\
University of California, Irvine\footnote{Email addresses: helen.meskhidze@uci.edu (Helen Meskhidze), weatherj@uci.edu (James Owen Weatherall) \\ H.M. framed the project and drafted the present paper. The statement and proof of the degeometrization theorem were a collaboration between H.M. and J.O.W.}
\\
\vspace{2em}
Draft of \today\\
\end{centering}

\begin{abstract}
Teleparallel gravity, an empirically equivalent counterpart to General Relativity, represents the influence of gravity using torsional forces. It raises questions about theory interpretation and underdetermination. To better understand the torsional forces of Teleparallel gravity, we consider a context in which forces are better understood: classical spacetimes. We propose a method of incorporating torsion into the classical spacetime context that yields a classical theory of gravity with a closed temporal metric and spacetime torsion. We then prove a result analogous to the Trautman degeometrization theorem, that every model of Newton-Cartan theory gives rise, non-uniquely, to a model of this theory.
\end{abstract}

\section{Introduction}

How should we understand gravitational influence?\footnote{Hours before submitting the, the authors became aware of a very recent paper on a similar topic \citep{TNCG}.  A cursory review suggests that the theory described in that paper differs from what we describe here, but since there is not sufficient time before the submission deadline, we leave an analysis of the relationship to future work.}   In traditional formulations of Newtonian Gravity (NG), gravitational influence is understood as a force. Gravitational force is mediated by a gravitational potential, which is itself related to the distribution of matter. This means that, in Newtonian Gravity, massive bodies exert attractive gravitational forces on one another. Our current best theory of gravity, General Relativity (GR), presents a different understanding of gravitational influence. GR is thought to have taught us that gravitational influence should be properly understood as a manifestation of spacetime curvature. In particular, massive bodies curve spacetime and gravitational influence is a manifestation of this curvature. This means that, unlike the flat spacetime background of traditional Newtonian Gravity, GR posits a curved spacetime that depends dynamically on the distribution of matter. 

Interestingly, the lessons of GR can be applied in the non-relativistic context.  It is possible to formulate a non-relativistic (i.e., classical) theory of gravity with curvature. This theory goes by the name of Newton-Cartan theory (NCT; sometimes referred to as ``geometrized Newtonian gravity'') and, while space is flat in this theory, space\emph{time} is curved. The curvature in NCT is dynamically determined by the matter distribution, and gravitational influences are a manifestation of the curvature of spacetime. Models of NCT are systematically related to models of GR as well as Newtonian Gravity. 

The above picture is complicated by the existence of a gravitational theory that is empirically equivalent to General Relativity, but represents gravitational influence as a force and is set on a flat spacetime background: Teleparallel Gravity (TPG). In contrast to Newtonian Gravity, the forces of TPG feature torsion (or, spacetime twisting). TPG raises both questions regarding underdetermination and more fundamental conceptual questions: Which theory, General Relativity or Teleparallel Gravity, describes our world?\footnote{This question has also been addressed by Eleanor Knox \citeyearpar{KnoxTeleparallel}. Our discussion here is influenced by her insights about TPG.} How should we understand the torsional forces posited by Teleparallel Gravity? And what is the relation between TPG and the other gravitational theories mentioned above?

Addressing these questions will be the goal here. We suggest that to better understand the forces of TPG, a natural place to begin is another gravitational theory that employs forces, namely, Newtonian Gravity. Newtonian Gravity, however, is a non-relativistic theory and gravitational force does not involve torsion. A classical theory of gravity with torsional forces will prove to be a more informative comparison. Developing such a theory to address the above questions will be the goal of this paper.   

Beyond the motivations already outlined, formulating a classical spacetime with torsion will also have implications for various projects in the physics literature. There have been proposals for torsional classical theories to describe the fractional quantum Hall effect \citep{Geracie+15} and to serve as the lower-dimensional reductions to 5D quantum gravity \citep{Christensen+14,Bergshoeff+14,Hartong+Obers15,Afshar+16,OFarrill20}. Though we will not discuss these projects in detail, we will discuss the classical spacetimes with torsion that they develop. In so doing, we will trouble one assumption they share and propose an alternative treatment of time. In particular, it is widely claimed that a classical spacetime with torsion cannot have a temporal metric that is closed, in the sense of differential forms.  As we discuss, this is not true---or rather, it holds only in the presence of a further condition that is motivated only by specific applications of classical spacetimes with torsion. 

To build a non-relativistic theory with torsion, we begin with some background on the theories mentioned above. We then discuss how to incorporate torsion in the non-relativistic context and what we would expect of such a theory in terms of how it represents space and time, as well as how it treats sources and forces. We next consider the relation of the proposed theory to other classical theories with torsion in the literature. Then finally state and prove a theorem, analogous to \citet{Trautman1965}'s degeometrization theorem, that establishes that, associated with every model of Newton-Cartan theory, there exists a (non-unique) model of Newtonian gravitation with torsion with the same mass density and particle trajectories.
 
\section{Background}

Let us begin by making the claims of the introduction more precise. As mentioned, our current best theory of gravity, General Relativity, represents gravitational influence through the curvature of spacetime. We take a model of that theory to be a pair, $(M, g_{ab})$ where $M$ is a smooth, connected, four-dimensional, paracompact, Hausdorff manifold, and $g_{ab}$ is a smooth, Lorentz-signature metric on $M$. As a relativistic theory, GR places an upper bound on the speed of light. This feature of GR is formalized by the metric, $g_{ab}$, which determines a light cone structure. Light-like particles (i.e., photons) follow trajectories along the cone while massive particles follow trajectories that lie inside the light cones. 

In contrast, Teleparallel Gravity is set on a flat spacetime background and represents gravitational influence through forces using torsion. Like GR, TPG is a relativistic theory (it posits a Lorentz-signature metric) and, as mentioned, TPG is empirically equivalent to GR, at least locally. We present the formal apparatus for understanding torsion below, but let us first develop an intuition for it. The torsion tensor characterizes the twisting of the tangent space as it is parallel transported along a curve. One can imagine parallel transporting two end-to-end vectors along one other. When the torsion is vanishing, this procedure yields a parallelogram. However, in spaces with torsion, the parallelograms break because the vectors do not end up tip-to-tip. As Cai, Capozziello, De Laurentis, and Saridakis put it: 

\begin{quotation}
...while in curved spaces considering two bits of geodesics and displacing one along the other will form an infinitesimal parallelogram, in twisted spaces the above procedure of displacing one geodesic bit along the other leads to a gap between the extremities, i.e. the infinitesimal parallelogram breaks. This implies that in performing the parallel transportation of a vector field in a space with torsion, an intrinsic length---related to torsion---appears. \citeyearpar[5]{Cai2016}
\end{quotation}  

A non-relativistic spacetime (i.e., that of Newton-Cartan theory and Newtonian gravity) can, in general, be expressed as $(M, t_a, h^{ab}, \tilde\nabla)$. $M$ is, as before, a smooth, connected four-dimensional, Hausdorff, paracompact manifold. The metric of GR, however, is replaced by two degenerate metrics: the temporal metric, $t_a$, and the spatial metric, $h^{ab}$. These metrics are orthogonal to one another (i.e., $h^{ab}t_b =0$). If the temporal length of a vector is non-vanishing, we characterize it as ``timelike'' (else, ``spacelike''). Finally, we require the derivative operator to be metric compatible ($\tilde\nabla_at_b=\mathbf{0}$ and $\tilde\nabla_ah^{bc} = 0$). These conditions ensure that the metric and affine structures agree.\footnote{In GR, the metric uniquely picks out a derivative operator and so it need not be explicitly specified. Since this is not the case in classical spacetimes, one needs to specify the derivative operator explicitly.}

\section{A classical gravitational theory with torsion}

We now describe the features of a gravitational theory set in (flat) spacetime with classical metrics and torsion.

\subsection{Torsion}

We begin with the most pressing issue: how to represent torsion in a classical spacetime. Since the displacement of vectors along one another is given by the derivative operator, to derive a theory of gravity with possibly non-vanishing torsion in the classical context, we must adjust the conditions on our derivative operator. 
We define a generic (i.e., not specific to the classical context) derivative operator following Malament's \citeyearpar{MalamentGR} conventions but drop the last requirement $(DO6)$---that the action of two derivative operators on a scalar field commute---to allow torsion. This leaves: 

\begin{definition} $\nabla$ is a (covariant) derivative operator on $M$ if it satisfies the following conditions: 

\begin{enumerate}[label=(D0\arabic*)]
 \item $\nabla$ commutes with addition on tensor fields.
 \item $\nabla$ satisfies the Leibniz rule with respect to tensor multiplication.
 \item $\nabla$ commutes with index substitution.
 \item $\nabla$ commutes with contraction.
 \item For all smooth scalar fields $\alpha$ and all smooth vector fields, $\xi^n$:
    \begin{equation*}
        \xi^n\nabla_n\alpha = \xi(a).
    \end{equation*}
\end{enumerate}
\end{definition}
Instead of requiring that $\nabla$ commute in its action on scalar fields $(D06)$, we set this to be the torsion tensor. 

\begin{definition}
Let $\nabla$ be a covariant derivative operator on the manifold $M$. Then, there exists a smooth tensor field $\tensor{T}{^a_{bc}}$, the torsion tensor, which is defined by: 
\begin{equation*}
    2\nabla_{[a}\nabla_{b]} \alpha = \nabla_a\nabla_b \alpha - \nabla_b\nabla_a \alpha = \tensor{T}{^c_{ab}} \nabla_c \alpha
\end{equation*}
for all smooth scalar fields $\alpha$.
\end{definition}


Note, from the definition above, that $T^a{}_{bc}$ is anti-symmetric in its lowered indices, $b$ and $c$:
\begin{align*} 
\tensor{T}{^a_{bc}} = - \tensor{T}{^a_{cb}} 
\end{align*}
As in the torsion-free case, the action of any two (possibly-torsional) derivative operators, $\nabla$ and $\tilde\nabla$, can be related by a smooth tensor field $\tensor{C}{^a_{bc}}$, with the property that for any smooth vector field $\tensor{\xi}{^a}$, $\nabla_a \tensor{\xi}{^b}=\tilde\nabla_a \tensor{\xi}{^b} - \tensor{C}{^b_{an}}\tensor{\xi}{^n}$ (and likewise for other tensor fields). In this case we write $\nabla=(\tilde\nabla,\tensor{C}{^a_{bc}})$  Unlike in the torsion-free case, however, this field $\tensor{C}{^a_{bc}}$ need not be symmetric in its lower indices.  Instead we have
\[
2\tensor{C}{^a_{[bc]}}=\tensor{T}{^a_{bc}}-\tensor{\tilde{T}}{^a_{bc}}.
\]

\subsection{Time and space}

We now consider how to represent time and space in a classical spacetime theory with torsion.  As in standard Newtonian gravitation, we will assume that spacetime has a temporal metric $t_a$ and spatial metric $h^{ab}$, both of which will be compatible with the possibly-torsional derivative operator $\nabla$. In standard models of Newtonian gravitation, without torsion, it follows from the compatibility of the temporal metric with the (torsion-free) derivative operator that $t_a$ is closed, i.e., $d_a t_b=0$, where $d$ is the exterior derivative. This implies that $t_a$ is locally exact, i.e., $t_a=\nabla_at$ for some smooth time function, $t$. Physically, the availability of a time function means that we can have a well-defined notion of the temporal distance between points. 
Indeed, if $M$ is simply connected, then we will have a global time function,  $t : M \rightarrow \mathbb{R} $. This means our spacetime consists of global simultaneity slices stacked through time and any two global time functions will differ only in their assignment of the zero-point for the time scale. Compatibility with a torsional derivative operator no longer implies that $t_a$ is closed in general.  However, we will assume $t_a$ is closed and we will only consider derivative operators compatible with closed temporal metrics in what follows.

The curvature of a spacetime is formalized by the Riemann curvature tensor. Intuitively, the Riemann tensor measures the degree to which a vector fails to return to its original value when parallel transported around a closed loop. More formally, it measures the degree to which the second covariant derivatives fail to commute. For a torsion-free spacetime, it is defined as
\begin{equation*}
R^a{}_{bcd} \xi^b = -2\tilde\nabla_{[c} \tilde\nabla_{d]}\xi^a.
\end{equation*}

In spaces with torsion, we adjust the definition of the Riemann tensor to include the contribution from the torsion tensor.\footnote{See \cite{JensenGRwithTorsion} for a derivation, though his sign conventions differ from ours.} This yields

\begin{equation*}
     R^a{}_{bcd}\xi^b = 
      - 2 \nabla_{[c} \nabla_{d]}\xi^a + T^n{}_{cd}\nabla_n\xi^a
\end{equation*}

There is a valuable formula relating the curvatures of two derivative operators with torsion.  If $\nabla=(\tilde\nabla,\tensor{C}{^a_{bc}})$, then
\[
\tensor{R}{^a_{bcd}}=\tensor{\tilde{R}}{^a_{bcd}}+2\tilde\nabla_{[c}\tensor{C}{^a_{d]b}}+ 2\tensor{C}{^p_{[c|b|}}\tensor{C}{^a_{d]p}} - \tensor{\tilde{T}}{^m_{cd}}
\tensor{C}{^a}_{mb}.
\]
Note only the torsion of $\tilde\nabla$ appears in this equation.

Let us compare the spatiotemporal geometry of NG and NCT. NG posits that space and time are both flat (i.e., ``spacetime is flat''), implying that the Riemann tensor, $R^a{}_{bcd}$, vanishes entirely. NCT, by contrast, only requires spatial flatness (i.e., ``space is flat''). We formalize this condition as $R^{abcd} = \mathbf{0}$, where indices are raised using $h^{ab}$, and interpret it as saying that the parallel transport of spacelike vectors in spacelike directions is, at least locally, path independent.\footnote{\label{curvatureConstraintNCT}Often, a stronger condition is adopted in NCT, that $R^{ab}{}_{cd} = \mathbf{0}$. $R^{ab}{}_{cd} = \mathbf{0}$ is equivalent to $R^{abcd} =0$ if and only if $R^{abcd} = \mathbf{0}$ \textit{and} there exists a local, unit timelike vector field $\xi^a$ that is rigid and twist-free (\citealt[Proposition 4.3.1]{MalamentGR}). Note that $R^{ab}{}_{cd} = \mathbf{0}$ implies that  $R^{abcd} = \mathbf{0}$ as we can simply raise the indices: $\mathbf{0} = R^{ab}{}_{pq}h^{pc}h^{qd} = R^{abcd}$).} 

To develop a theory most like Teleparallel Gravity in the classical context, we will require the curvature of our spacetime to vanish, $R^a{}_{bcd}$.
This is because TPG is set on a flat spacetime background, and we are seeking a classical theory analogous to it.  

We have not, thus far, placed any constraints on the torsion tensor. Recall that in NCT, the spatial curvature vanishes. Analogously, we propose that the spatial torsion of our spacetime vanish (i.e., $T^{abc} = 0$). The vanishing of the spatial torsion will yield a theory like NCT but with torsion, not curvature, encoding gravitational influence.

\subsection{Sources and forces}
\label{sec:sourcesforces}
Finally, let us consider how we expect sources to exert (torsional) force in our theory. It will be instructive to consider the treatment of sources and forces in the non-torsional, classical context first. In Newtonian Gravity, bodies are subject to gravitational forces and force is mediated by a gravitational potential ($\phi$). The four-velocity, $\xi^a$, of a particle satisfies
 
\begin{align}
\label{eq:ForceEqnNG}
    -\nabla_a \phi = \xi^n \nabla{}_n \xi_a, 
\end{align}

where $\phi$ is a smooth, scalar field and $\nabla$ denotes the flat, torsion-free derivative operator of standard NG. The right-hand side of the equation describes the acceleration that the test point particle undergoes in the presence of the gravitational potential, $\phi$. The gravitational potential further satisfies Poisson's equation, relating it to the distribution of matter 

\begin{align}
\label{eq:PoissonEqnNG}
    \nabla_a \nabla{}^a \phi = 4 \pi \rho,
\end{align}
 
where $\rho$ is the Newtonian mass density function. 

In Newton-Cartan theory, like in GR, the curvature of spacetime means that inertial motion is governed by the geodesic principle: in the absence of external (non-gravitational) forces, bodies move along the geodesics of (curved) spacetime. The equation of motion is given as 

\begin{align}
    \xi^n\tilde\nabla_n\xi^a = \mathbf{0},
\end{align}
where $\tilde\nabla$ is the curved derivative operator of NCT. 

To account for spatiotemporal curvature, NCT adopts a geometrized form of Poisson's equation, relating the distribution of matter to the curvature of spacetime, 

\begin{align}
\label{eq:PoissonEqnNCT}
    R_{ab} = 4\pi\rho t_a t_b.
\end{align}

As it turns out, models of NG and NCT are systematically related. The Trautman geometrization lemma and degeometrization theorem describe these relations. Let us consider the recovery of models of NG from NCT. This is the direction in which force terms arise and so it will be instructive in formulating torsional forces. To build up to the degeometrization theorem, we will first consider the derivative operators of each theory. One can show that in the non-torsional context, one has the following result. 

\begin{proposition}  \citep[Proposition 4.1.3]{MalamentGR} Let $(M, t_a, h^{ab}, \tilde\nabla)$ be a classical spacetime. Let $\nabla= (\tilde\nabla, C^a{}_{bc})$ be a second derivative operator on $M$. Then, $\nabla$ is compatible with $t_a$ and $h^{ab}$ if and only if $C^a{}_{bc}$ is of the form: 
\begin{align*}
C^a{}_{bc} = 2h^{an}t_{(b}\kappa_{c)n}    
\end{align*}
where $\kappa_{ab}$ is a smooth anti-symmetric field on $M$ and the parentheses denote symmetrization. 
\end{proposition}

If we permit derivative operators with torsion, a broader class of derivative operators are compatible with the classical metrics. We now have the following generalization of the preceding proposition.

\begin{proposition}\label{compatibility}
Let $(M, t_a, h^{ab}, \tilde\nabla )$ be a classical spacetime with (possibly) non-vanishing torsion. Let $\nabla{} = (\tilde\nabla, C^a{}_{bc})$ be another derivative operator on $M$ also with (possibly) non-vanishing torsion (i.e., $2C^a{}_{[bc]} = T^a{}_{bc} -\tilde{T}{}^a{}_{bc}$). Then $\nabla$ is compatible with $t_a$ and $h^{ab}$ if and only if $C^a{}_{bc}$ is of the form:
\begin{align*}
    \tensor{C}{^{a}_{bc}} = 2 h^{ar}\kappa_{[r|b|c]}.
\end{align*}
If, in addition, we require $\tilde{T}^{abc}=T^{abc}=\mathbf{0}$, then
\begin{align*}
    \tensor{C}{^{a}_{bc}} = 2 h^{ar}(x_{rb}t_c +y_{rc}t_b).
\end{align*}
where $x_{ab}$ is an arbitrary smooth tensor field and $y_{rc}$ is any antisymmetric field.
\end{proposition}

As we can see, in the presence of torsion, there is considerable freedom to define metric-compatible derivative operators.  Below, we will limit attention to flat, metric-compatible derivative operators whose torsion has the form $T^a{}_{bc} = 2 F^a{}_{[b}t_{c]}$ where $F^a{}_{b}$ is a smooth rank (1,1) tensor field, spacelike in the $a$ index.  This is tantamount to stipulating that $y_{ab}$ in the Prop. \ref{compatibility} vanishes. This restriction clearly satisfies the above-outlined general form and ensures that the spatial torsion vanishes. Furthermore, as will be seen in the below theorem, we can recover the standard, torsion-free connecting field assumed for the degeometrization theorem as a special case of the above.

In describing the difference between the derivative operators, the connecting field is closely related to the force field that arises in the degeometrization of a model of Newton-Cartan theory.  In the torsion-free context, one typically assumes a connecting field of the form $C^a{}_{bc} = t_b t_c \tilde\nabla{}^a \phi$.\footnote{A connecting field of this form satisfies the more general constraint for the connecting field between the derivative operators of any two classical spacetime models if we take $\kappa_{cn}$ from above to be $t_{[c}\tilde\nabla_{n]} \phi$.  That some $\phi$ exists with the necessary properties to make this derivative operator flat depends on several background assumptions that we suppress for reasons of space.} The force term is then just the contracted connecting field: $C^a{}_{rn}\xi^{r}\xi^n = \tilde\nabla^a \phi$. 

To adapt this to the torsional context, we want to consider the connecting field relating a non-torsional, flat derivative operator to a torsional one. We will capture the impact of the torsion on the trajectories of test bodies using the above-mentioned tensor field, $F^a{}_b$. In other words, we want $F^a{}_b$ to play the role of a torsional force term. Given a timelike geodesic of $\tilde{\nabla}$ with unit tangent field $\xi^a$, the force equation we expect to be satisfied is
\begin{align}
\label{eq: modifiedForceEq}
\xi^n {\nabla}{}_n\xi^a = \xi^n\tilde\nabla_n\xi^a - C^a{}_{rn}\xi^r \xi^n = -F^a{}_n t_c \xi^n\xi^c = - F^a{}_n \xi^n, 
\end{align}
where $\nabla$ is the derivative operator of our torsional spacetime. 

Finally, we want to relate the torsional force term to gravitational sources. In other words, we want to formulate a field equation that is the torsional analog to Poisson's Equation. It will turn out to be 
\begin{align}\label{torsionalPoisson}
    \delta^n_a \nabla_{[n}F^a{}_{b]}= 2 \pi \rho t_b.
\end{align}

Again, Poisson's equation will be recovered as a special case of Eq. \eqref{torsionalPoisson}, but Eq. \eqref{torsionalPoisson} more generally establishes a relation between the first-derivative of the force term and the mass distribution along the temporal direction. 

\subsection{Comparison to other classical theories with torsion}

As mentioned above ($\S$1), a small literature has recently emerged in physics surrounding torsional classical spacetime theories. Many in this literature are interested in incorporating torsion in the classical context to address the different notions of time proposed by GR and Quantum Gravity (QG): though GR does not admit a global notion of simultaneity, in some formulations, QG does. To resolve this difference, some authors have proposed taking the notion of time presented in QG as fundamental and allowing relativistic time to emerge at large distances. Then, motivated by the holographic principle (i.e., that a volume of space can be thought of as encoded in the lower dimensional boundary of that volume),\footnote{The projects in this literature are also sometimes motivated as attempts to find further holographic correspondences beyond the AdS/CFT correspondence (see \citet[1]{Christensen+14}).} this literature considers 5D QG and its 4D reduction. The holography considered is between Ho\u{r}ava–Lifshitz gravity and a new theory of classical gravity: twistless torsional Newton-Cartan (TTNC) theory.\footnote{The first paper developing this theory was \cite{Christensen+14First}. A slew of others followed including \citet{Christensen+14,Bergshoeff+14,Hartong+Obers15,Afshar+16,OFarrill20}.} 

As noted above, typically in the classical spacetime context, we take $t_a$ to be closed (i.e., $d_at_b = 0$), which means it is locally exact and determines a local time function. This is a consequence of its compatibility with any torsion-free derivative operator. We also adopt this assumption in the torsional gravitational theory developed here. 

The TTNC formalism, by contrast, starts with NC theory but claims that \textit{taking $\partial_\mu t_\nu =0$}, where $\partial$ is a (torsion-free) coordinate derivative operator \textit{will always result in a torsion-free spacetime.}  Thus they do not require temporal metrics to be compatible with any torsion-free derivative operator; more generally, they do not require that $t_a$ is closed. Consider the following characteristic passage: 

\begin{quotation}
The absence of torsion implies that the temporal vielbein\footnote{It is common to see formulations of classical gravity with torsion presented with the vielbein formalism typical of presentations of TPG. One can simply think of the temporal vielbein here as the temporal metric.} $\tau_\mu$ corresponds to a closed one-form and that it can be used to define an absolute time in the space–time...TTNC geometry is characterized by the fact that the temporal vielbein is hypersurface orthogonal but not necessarily closed. \cite[3]{Bergshoeff+14}
\end{quotation}

\noindent In order to derive a hypersurface orthogonal temporal metric, such authors appeal to Frobenius' theorem. This allows them to argue that a spacetime admits a foliation with a time flow orthogonal to the Riemannian spacelike slices if and only if it satisfies the hypersurface orthogonality condition (i.e., $t_{[a} \partial_b t_{c]} = 0$). Notably, the ``hypersurface orthogonality condition'' is a weaker condition than the condition that the temporal metric be closed. 

A series of questions emerge from the above discussion: Why does the TTNC literature claim that closed temporal metrics, and metric compatibility more generally, are in tension with torsion? And how does the theory described above incorporate torsion and a closed temporal metric, and thus a notion of absolute time?  The answers lies in the form of the connection assumed by the TTNC literature.

Geracie and collaborators define a spacetime derivative operator $\nabla=(\partial,\Gamma^a{}_{bc})$, where they require the form of the connecting field $\Gamma^a{}_{bc}$ to be
\begin{align*}
    \Gamma^a{}_{bc} = v^a \partial_b t_c + \frac{1}{2} h^{an} (\partial_b \hat{h}_{cn} + \partial_{c}\hat{h}_{b n} - \partial_n \hat{h}_{bc}), 
\end{align*}
where $v^a$ is a unit timelike field, and $\hat{h}_{ab}$ is a spatial projection field determined by $v^a$, such that $h^{an}\hat{h}_{nb}=\delta^a{}_b-v^at_b$ \citeyearpar[Eq. 77]{Geracie+15}.\footnote{We have translated their formalism to match our notation.}  This definition is motivated by the standard definition of a Levi-Civita derivative operator, and the terms in the parentheses are always symmetric in $b,c$.  It follows that the torsion is given by $T^a{}_{bc} = 2 \Gamma^{a}{}_{[bc]} =  2v^a \partial_{[b}t_{c]}$ (see, e.g., \citet[Eq. 79]{Geracie+15}). (Indeed, the name ``twistless torsional NCT,'' then, comes from the fact that torsion vanishes on spacelike hypersurfaces but not in general.)  

And so it is true that if $t_a$ is closed, the torsion of this derivative operator would vanish.  However, \textit{this is only because they have adopted such a strict definition for their connection}. Put differently, their connection ensures that the only way to allow torsion is to sacrifice having a closed temporal metric. But there are many other torsional derivative operators that are compatible with a closed temporal metric.  Once we allow for a broader class of connections, as is done in the present paper, we recover metric compatibility and a notion of absolute time. 

Classical spacetimes with torsion have also been the subject of philosophical analysis. In particular, James Read and Nicholas Teh \citeyearpar{ReadTeh} have developed a method for ``teleparallelizing'' in the classical context. A central aim of their project is to show the relation between this classical theory and its relativistic counterpart, TPG. They begin with NCT and teleparallelize to construct a classical spacetime with torsion. Their procedure involves a ``mass torsion'' term that plays the role of a force. Contrary to the results derived in the present paper, they claim their teleparallelization method yields standard Newtonian Gravity. Their proposal is interesting in its own right. However, we would argue that the spacetime developed in the present paper, in so far as it features \textit{spacetime} torsion instead of mass torsion, is a stronger analog to a classical TPG.

\section{Degeometrization with Torsion}
We now state and prove a theorem analogous to the Trautman degeometrization theorem.  This result establishes that for every model of Newton-Cartan theory, there is a corresponding model of the classical analog to teleparallel gravity described above. 

\begin{theorem}
Let $(M,t_a,h^{ab},\tilde{\nabla})$ be a classical spacetime (without torsion) satisfying:

\begin{align}
\tilde{R}_{ab} &= 4\pi\rho t_b t_c
\end{align}
\begin{align}
\tilde{R}^{ab}{}_{cd} &= \mathbf{0}
\end{align}

for some smooth scalar field $\rho$.  Then given any point $p$ in $M$, there is an open set $O$ containing $p$ and a pair $(\nabla,F^a{}_{b})$ on $O$, where $\nabla$ is a derivative operator and $F^a{}_b$ is a smooth rank (1,1) tensor field, which together satisfy the following conditions:
\begin{enumerate}
\item $\nabla$ is compatible with $t_a$ and $h^{ab}$;
\item $\nabla$ is flat;
\item $\nabla$ has torsion $T^a{}_{bc}=2F^a{}_{[b}t_{c]}$;
\item For all timelike curves with unit tangent field $\xi^a$, $\xi^n\tilde{\nabla}_n\xi^a=\mathbf{0}$ if and only if $\xi^n\nabla_n\xi^a = -F^a{}_n\xi^n$; and
\item $(\nabla_a,F^a{}_b)$ together satisfy the field equations $\delta^n{}_a\nabla_{[n}F^a{}_{b]}=2\pi\rho t_b$.
\end{enumerate}

The pair $(\nabla,F^{a}{}_b)$ is not unique.  Moreover, there exist pairs $(\nabla,F^a{}_b)$, satisfying the conditions above, for which the torsion is non-vanishing. 
\end{theorem}

Proof. Existence follows from the Trautman degeometrization theorem \citep[Proposition 4.2.5]{MalamentGR}. Fix any classical spacetime $(M,t_a,h^{ab},\tilde{\nabla})$ satsifying $\tilde{R}{}^{ab}{}_{cd}=\mathbf{0}$ and $\tilde{R}_{ab}=4\pi\rho t_a t_b$ for some smooth scalar field $\rho$.  Choose a point $p$ and a rigid and twist-free field $\eta^a$ defined on some neighborhood of $p$, and let $\varphi^a = \eta^n\tilde{\nabla}_n\eta^a$ be the acceleration field associated with $\eta^a$.  Then the pair $(\nabla, F^a{}_b)$, where $F^a{}_b=\varphi^at_b$ and $\nabla=(\tilde{\nabla},F^a{}_bt_c)$, satisfies conditions 1-5, with torsion $T^a{}_{bc}=2\varphi^at_{[b}t_{c]}=\mathbf{0}$, by arguments given in Malament's proof.  Indeed, in this case the field equation $\delta^n{}_a\nabla_{[n}F^a{}_{b]}=2\pi t_b$ reduces to
\[
2\pi t_b =\frac{1}{2}\delta^n{}_a\left(\nabla_{n}\varphi^at_b - \nabla_b \varphi^a t_n\right) = \frac{1}{2}t_b\nabla_a\varphi^a
\]
and the resulting structure is a model of ordinary Newtonian gravitation with gravitational field $\varphi^a$.  (If one assumed further that $\tilde{R}{}^a{}_b{}^c{}_d = \tilde{R}{}^c{}_d^a{}_b$, one could conclude that $\varphi^a=\nabla^a\varphi$ for some smooth scalar field $\varphi$, possibly on a subneighborhood of $O$.)

Non-uniqueness also follows from the Trautman degeometrization theorem. We wish to show, however, that there exist pairs $(\nabla, F^a{}_b)$ satisfying conditions 1-5 with non-vanishing torsion.  We do so by direction construction. Let $\nabla=(\tilde{\nabla},\varphi^at_bt_c)$ be the flat derivative operator (without torsion) considered above.   Choose any spacelike vector $x^a$ at $p$, and extend it to a neighborhood of $p$ by parallel transport via $\check{\nabla}$.  Finally let $\psi$ be any smooth scalar field defined near $p$ whose gradient is spacelike and normal to $x^a$.  Now define $\hat{F}^a{}_b = x^a\nabla_b\psi$ and $\check{F}{}^a{}_{b}=\varphi^a t_b+\hat{F}{}^a{}_b$.  Then the pair $(\check{\nabla},\check{F}{}^a{}_b)$, where $\check{\nabla}=(\nabla,\hat{F}{}^a{}_bt_c)$, satisfies conditions 1-5 with torsion $T^a{}_{bc}=2x^at_{[c}\nabla_{b]}\psi\neq 0$.

To see that 1 is satisfied, observe that $\check{\nabla}_at_b = t_n \check{F}{}^n{}_at_b = 0$; and $\check{\nabla}_a h^{bc}= F^b{}_at_nh^{nc} + F^c{}_at_nh^{bn}$. For 2, note that since $\nabla$ is flat and torsion-free, and $x^a$ is constant with respect to $\nabla$, we have
\[
\check{R}{}^a{}_{bcd} = 2x^at_b\nabla_{[c}\nabla_{d]}\psi + 2x^pt_px^at_b\nabla_{[c}\psi \nabla_{d]}\psi =\mathbf{0}
\]
where the first term vanishes because $\nabla$ is torsion-free and the second because $x^a$ is spacelike.  3 follows from the definition of $\check{\nabla}$ and the fact that $\nabla$ is torsion-free. 4 follows because for all unit timelike vector fields $\xi^a$,
\begin{align*}
\xi^n\tilde{\nabla}_n\xi^a = \mathbf{0} &\Leftrightarrow& \xi^n\nabla_n\xi^a = -\nabla^a &\Leftrightarrow& \xi^n\check{\nabla}_n\xi^a = -\varphi^a - \hat{F}{}^a{}_n\xi^n =  -F^a{}_n\xi^n.
\end{align*}
Finally, 5 is satisfied because
\begin{align*}
\delta^n{}_a\check{\nabla}_{[n}F^a{}_{b]}&=\delta^n{}_a\check{\nabla}_{[n}(\varphi^at_{b]}+x^a\nabla_{b]}\psi)\\
&=2\pi\rho t_b+\delta^n{}_a\check{\nabla}_{[n}x^a\check{\nabla}_{b]}\psi)\\
&=2\pi\rho t_b + x^n\check{\nabla}_{[n}\check{\nabla}_{b]}\psi\\
&=2\pi\rho t_b + \frac{1}{2}x^nT^a{}_{nb}\check{\nabla}_a\psi\\
&=2\pi\rho t_b + x^nx^at_{[b}\nabla_{n]}\psi\check{\nabla}_a\psi\\
 &=2\pi\rho t_b
 \end{align*}
where in the first equality we use the facts that $0=\nabla_a x^b = \check{\nabla}_a x^b - \hat{F}{}^b{}_at_bx^b = \check{\nabla}_a x^b$ and that $\nabla$ and $\check{\nabla}$ agree on scalar fields (because all derivative operators do); while in the final equality we use the fact that $\check{\nabla}_a\psi$ is normal to $x^a$. \hspace{.25in}$\square$

\section{Discussion}

The general proof strategy is to leverage the original Trautman degeometrization theorem results. We show that NG can be recovered as a special (torsion-free) case of the theorem presented above. By broadening the class of allowed derivative operators, the non-uniqueness results establish the possibility of a classical spacetime with non-vanishing torsion. 

There are some important differences between our result and the Trautman theorem.  We do not give necessary and sufficient conditions to construct new pairs $(\nabla,F^a{}_b)$ from old ones satisfying 1-5. 
 This is because, unlike the situation with vanishing torsion, the derivative operators associated with models of the torsional theory (for some model of NCT) do not appear to form an affine space. Nonetheless, we are able to establish the non-uniqueness of torsional models, and we give a general strategy for constructing alternative models with torsion associated with a given model of NCT. It would be interesting to provide a complete description of this space.   

We also do not require our model of Newton-Cartan theory to satisfy $R^a{}_b{}^c{}_d=R^c{}_d{}^a{}_b$, as the Trautman theorem does.  This is because the role of that condition is to ensure that a certain field $\varphi^a$ is closed, and therefore locally exact.  We do not invoke that field in the result, and so we drop the condition.  In that sense, we generalize the Trautman theorem. Finally, we note that more general versions of the theory (and theorem) discussed here are almost certainly possible.  For instance, one might consider force fields for which $y_{ab}$ from Prop. \ref{compatibility} is non-vanishing, among other variations \citep[c.f. \S4.5]{MalamentGR}.

\bibliography{main.bbl}
\end{document}